\documentclass[aps,floatfix,nofootinbib,preprintnumbers,amsmath, amssymb,showkeys,preprint]{revtex4-1}

\usepackage{epsf,epsfig,subfigure,graphicx,amsmath,amssymb}
\usepackage{color}
\usepackage{slashed}
\usepackage{hyperref}
\usepackage[normalem]{ulem}

\def\lsim{\mathrel{\rlap{\lower4pt\hbox{\hskip1pt$\sim$}}
    \raise1pt\hbox{$<$}}}         
\def\gsim{\mathrel{\rlap{\lower4pt\hbox{\hskip1pt$\sim$}}
    \raise1pt\hbox{$>$}}}         

\newcommand{\red}{\textcolor{black}}

\newcommand{\be}{\begin{equation}}
\newcommand{\ee}{\end{equation}}
\newcommand{\bea}{\begin{eqnarray}}
\newcommand{\eea}{\end{eqnarray}}

\def\circa#1{\,\raise.3ex\hbox{$#1$\kern-.75em\lower1ex\hbox{$\sim$}}\,}

\begin{document}
\title{\bf Indirect signature of dark matter \\
with the diphoton resonance at 750 GeV }

\author{
Jong-Chul Park}
\email{jcpark@cnu.ac.kr}
\affiliation{
Department of Physics, Chungnam National University, Daejeon 34134, Korea
}
\author{Seong Chan Park}
\email{sc.park@yonsei.ac.kr}
\affiliation{
Department of Physics \& IPAP, Yonsei University, Seoul 03722, Korea\\
Korea Institute for Advanced Study (KIAS), Seoul 02455, Korea
}

\vspace{2.0cm}

\begin{abstract}
Motivated by the recently reported diphoton resonance at $750$ GeV, we study a new axion-like bosonic portal model of dark matter physics.
When the resonance particle is identified as the pseudo-scalar mediator, via which the standard model sector would interact with the dark matter sector, the data from collider physics would provide profound implications to dark matter phenomenology.
In this paper, we first identify the preferred parameter space of the suggested portal model from the results of the LHC run with $\sqrt{s}=13$ TeV, and then we examine the dark matter signature taking into account the data from cosmic-ray experiments including Fermi-LAT dwarf galaxy $\gamma$-ray search, HESS $\gamma$-line search, and future CTA diffuse $\gamma$-ray and $\gamma$-line searches.
\end{abstract}


\maketitle

\section{Introduction}

\red{The ATLAS~\cite{ATLAS15} and CMS~\cite{CMS15} Collaborations reported a new resonance at around $750~ {\rm GeV}$ in the diphoton channel seen in the first data obtained at the $\sqrt{s}=13$ TeV LHC run in December 2015. In March 2016, the results were updated with more data and new analysis with different assumptions on the width and spin state of the resonance particle \cite{ATLAS16, CMS16}.  Interestingly, the claimed local significance stays high or even slightly grows from $3.9\sigma/2.6\sigma$ (ATLAS with 45 GeV decay width/CMS, December 2015) to  $3.9\sigma/2.9\sigma$ (ATLAS/CMS with 13 TeV data, March 2016). When (8+13 TeV) data are considered by CMS the significance becomes 3.4$\sigma$.  The global significance is still low($\lsim 2 \sigma$) for both experiments when the look-elsewhere effect is taken into account. The ATLAS result shows a slightly better fit to the data with a largish width $\Gamma/M \approx 6\%$ than with a narrow width $\Gamma/M\ll 1\%$ but the CMS result is consistent with narrow width approximation so that the result is not conclusive yet.
The resonance signals do not seem accompanied by significant missing energy, leptons or jets. The situation resembles the situation in 2011--2012 when the excess in diphoton channel was announced in the Higgs boson search.}

Since we also have observed that several $2\sigma \sim 3 \sigma$ `excesses'  disappeared as statistical fluctuations in the past,
we are extremely cautious in taking these observations as a signal of new physics beyond the standard model.
However, we think that the following points make the observation more interesting and many authors immediately suggested their interpretations:
\begin{itemize}
\item Both the ATLAS and CMS Collaborations reported the excess with $> 2\sigma$ significance, which is rather exceptional.\footnote{Indeed, the current situation is similar to the one when the announcement of the Higgs signature was made late 2011 with similar statistical significances from both experiments~\cite{ATLAS11, CMS11}.} \red{Even more interestingly, the significance grows with more data, which would strengthen our confidence. }
\item The locations of the reported excess are well within the experimental uncertainties:  $M_{\rm ATLAS} \approx 750$ GeV and $M_{\rm CMS} \approx 760$ GeV.
    We take 750 GeV as a representative value in our analysis below.
\item The excesses locate a bit far from the end points of the data from both experiments.
\item The excesses consist of 3-5 successive bins in the data from both experiments.
\end{itemize}

\begin{figure}
\begin{center}
\includegraphics[width=0.80\textwidth]{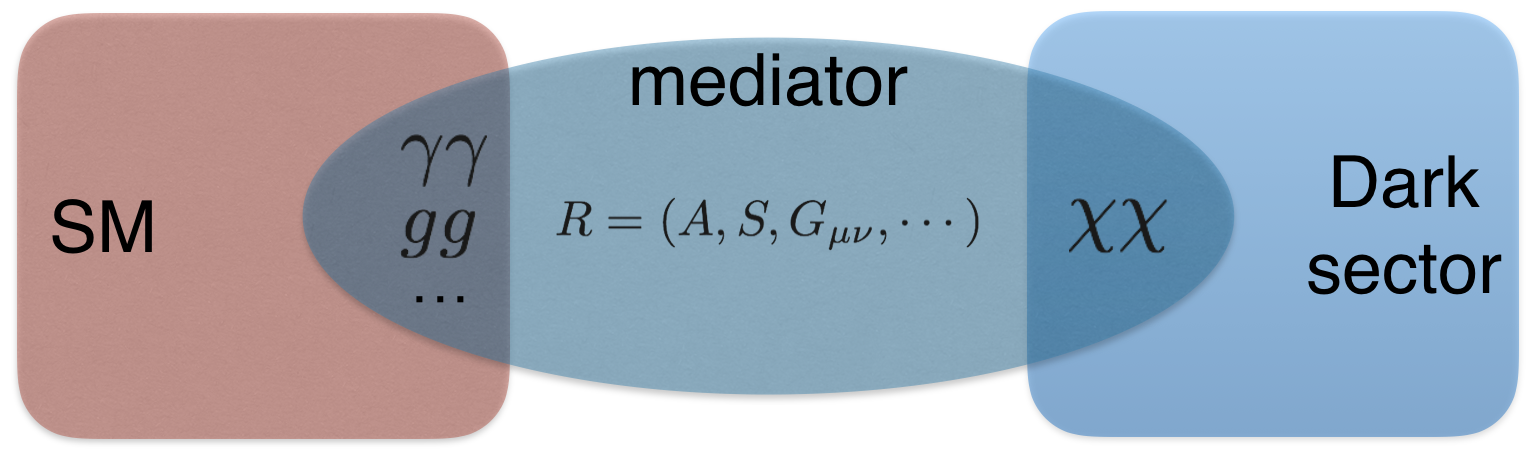}
\end{center}
\vspace*{-0.7cm}
\caption{ The schematic diagram of the standard model (SM)-bosonic mediator (Mediator) - dark sector.
The bosonic mediator can be a scalar ($S$), pseudo-scalar or axion ($A$), and spin-2 mediator (`graviton', $G$) with the mass of 750 GeV.}
\label{fig:diagram}
\end{figure}

Instead simply adding additional way of interpretation, we want to discuss a bit different phenomenological aspect of the current observation by connecting dark matter problem in this paper.
We consider new portal interactions between the dark matter sector and the standard model sector through the observed resonance state as the mediator. See Fig.~\ref{fig:diagram} which depicts the schematic diagram showing how the two sectors are linked via new portal interactions through scalar ($S$), pseudo-scalar ($A$) or tensor ($G$), which may be identified with the observed resonance at 750 GeV.
\footnote{While we were finishing our paper, we noticed some papers appeared on arXiv with different but related approaches to dark matter physics~\cite{dm1,dm2,dm3,dm4,dm5,dm6,dm7,dm8}.
We also see more recent works~\cite{dm9,dm10,dm11,dm12,dm13,dm14} in a related direction.}
We note that the most interesting case for dark matter indirect detection is found with the pseudo-scalar mediator, $A$, since other cases with $S$ and $G$ are all suppressed by the small velocity as $v^2\ll 1$ in current universe.
Thus, we will focus on the pseudo-scalar case in below and examine the details of indirect detection of the dark matter signature in cosmic-ray.

%
%

The content of the paper consists as follows.
In the next section, we will define our setup to discuss the 750 GeV mediator, and discuss collider signature and bounds from the LHC run-1 with $\sqrt{s}=8$ TeV as well as the new data with $\sqrt{s}=13$ TeV in the following section.
We examine the preferred parameter space from the collider physics data and found the potential detection of the indirect signature of dark matter in section III then conclude in section IV.

\section{A Pseudo-scalar portal
\label{sec:model}}

An intriguing possibility is that dark matter would communicate with the visible sector of the standard model (SM) via a `portal'.
In literature, various portals have been suggested: the Higgs portal ($\sim |H|^2{\cal O}_b$) and the neutrino portal via Yukawa interactions ($\sim \overline{\ell}H {\cal O}_f$), where ${\cal O}_{b}$ and ${\cal O}_{f}$ are bosonic and fermionic singlet operators of the dark sector, respectively.
In principle, however, the dark sector gauge symmetry structure can be extensively involved and may play a role to make the dark matter stable.~\cite{Huh:2007zw}

Here we take the resonance is nothing but the mediator between the standard model sector and the dark sector where dark matter belongs to.
The resonance, located at $750$ GeV, is bosonic with $s=0$ or $2$.
Because of Landau-Yang theorem (or Furry's theorem)~\cite{Landau}, a massive vector mediator is excluded.
The singlet pseudo-scalar $A$ interacts with the SM gauge bosons as well as a Dirac fermion dark matter $\chi$, described by the following effective Lagrangian,
\bea
{\cal L}_A=-\frac{A}{\Lambda} \Big(a_1 F^Y_{\mu\nu}{\tilde F}^{ Y\mu\nu}+a_2 W_{\mu\nu} {\tilde W}^{\mu\nu}+a_3 G_{\mu\nu} {\tilde G}^{\mu\nu}  \Big) - i\lambda_\chi A \,{\bar \chi}\gamma^5 \chi\,,
\eea
where the dual field strength tensor is ${\tilde F}_{\mu\nu}\equiv \frac{1}{2}\epsilon_{\mu\nu\rho\sigma} F^{\rho\sigma}$, etc.
For more generic Lagrangian with generic spin (i.e. $s=0,1,2$) and CP-states which couple to dibosons, see e.g. Ref.~\cite{kklp}.

In this model, the pseudo-scalar can be produced via gluon fusion at the LHC, decaying into a pair of SM gauge bosons.\footnote{See Ref.~\cite{Jaeckel:2012yz} for various constraints on pseudo-scalars coupled to two photons and gluons.}
When the pseudo-scalar decays substantially into $\gamma\gamma$ compared to $WW/ZZ/Z\gamma$, it can explain the diphoton excess recently reported.
If $m_A>2m_\chi$, the pseudo-scalar can also decay into a pair of dark matter leading to a signal with a large missing energy. This may help to understand a largish decay width $\Gamma/M \simeq 6\%$ as pointed out in Refs.~\cite{dm1,dm2,dm3,dm9} even though a narrow width is also allowed.

The similar pseudo-scalar resonance can play a role of mediator between the SM and dark matter sectors~\cite{lpp, Kyae:2013qna}.
The DM annihilation cross sections into gauge bosons are given by
\bea
\langle\sigma v_{\rm rel}\rangle_{\gamma\gamma} &=& \frac{\lambda^2_\chi a^2_{\gamma\gamma}}{4\pi \Lambda^2}\, \frac{16m^4_\chi}{(4m^2_\chi-m^2_A)^2+\Gamma^2_A m^2_A} + \mathcal{O}(v^2_{\rm rel}) \,, \\
\langle\sigma v_{\rm rel}\rangle_{Z\gamma} &=&
 \frac{\lambda^2_\chi a^2_{Z\gamma}}{8\pi\Lambda^2}\, \frac{16m^4_\chi}{(4m^2_\chi-m^2_A)^2+\Gamma^2_A m^2_A}\, \left( 1-\frac{m^2_Z}{4m^2_\chi} \right)^3 + \mathcal{O}(v^2_{\rm rel}) \,, \\
 \langle \sigma v_{\rm rel}\rangle_{ZZ} &=&  \frac{\lambda^2_\chi a^2_{ZZ}}{4\pi\Lambda^2}\, \frac{16m^4_\chi}{(4m^2_\chi-m^2_A)^2+\Gamma^2_A m^2_A}\, \left(1-\frac{m^2_Z}{m^2_\chi} \right)^{3/2} + \mathcal{O}(v^2_{\rm rel}) \,, \\
 \langle \sigma v_{\rm rel}\rangle_{WW} &=&  \frac{\lambda^2_\chi a^2_{WW}}{8\pi \Lambda^2}\, \frac{16m^4_\chi}{(4m^2_\chi-m^2_A)^2+\Gamma^2_A m^2_A}\, \left(1-\frac{m^2_W}{m^2_\chi} \right)^{3/2} + \mathcal{O}(v^2_{\rm rel}) \,, \\
 \langle \sigma v_{\rm rel}\rangle_{gg} &=&  \frac{2\lambda^2_\chi a^2_{gg}}{\pi \Lambda^2}\,\frac{16m^4_\chi}{(4m^2_\chi-m^2_A)^2+\Gamma^2_A m^2_A} + \mathcal{O}(v^2_{\rm rel})\,,
\eea
where
\bea
a_{\gamma\gamma}&=& a_1 \cos^2\theta_W + a_2 \sin^2\theta_W\,, \label{arr} \\
a_{Z\gamma}&=& (a_2-a_1) \sin(2\theta_W)\,, \label{aZr} \\
a_{ZZ} &=& a_1\sin^2\theta_W+ a_2 \cos^2\theta_W\,, \label{aZZ} \\
a_{WW} &=& 2a_2\,, \label{aWW} \\
a_{gg} &=& a_3 \label{agg} \,.
\eea
We note that all the gauge boson channels are $s$-wave.
Here, the partial decay rates of the pseudo-scalar are
\bea
\Gamma_A(\gamma\gamma)&=& \frac{m^3_A}{4\pi \Lambda^2} \,a^2_{\gamma\gamma}\,, \\
\Gamma_A(Z\gamma) &=& \frac{m^3_A}{8\pi \Lambda^2}\,a^2_{Z\gamma} \Big(1-\frac{m^2_Z}{m^2_A}\Big)^3\,, \\
\Gamma_A(ZZ) &=&  \frac{m^3_A}{4\pi \Lambda^2}\,a^2_{ZZ} \Big(1-\frac{4m^2_Z}{m^2_A}\Big)^{\tfrac{3}{2}}\,, \\
\Gamma_A(WW)&=&  \frac{m^3_A}{8\pi\Lambda^2} \,a^2_{WW} \Big(1-\frac{4m^2_W}{m^2_A}\Big)^{\tfrac{3}{2}}\,, \\
\Gamma_A(gg)&=&  \frac{2m^3_A}{\pi \Lambda^2}\, a^2_{gg}\,, \\
\Gamma_A({\bar\chi}\chi)&=& \frac{\lambda^2_\chi m_A}{8\pi} \Big(1-\frac{4m^2_\chi}{m^2_A}\Big)^{\tfrac{3}{2}} \theta(\frac{m_A}{2m_\chi}-1)\,. \label{Gamma2DM}
\eea

For $m_\chi > m_A$, dark matter can annihilate into a pair of pseudo-scalars with the annihilation cross section given by
\bea
\langle\sigma v_{\rm rel}\rangle_{AA}&=& \frac{\lambda^4_\chi}{24\pi} \frac{m^6_\chi}{(m^2_a-2m^2_\chi)^4}\, \Big(1-\frac{m^2_A}{m^2_\chi}\Big)^{5/2}\, v^2_{\rm rel}\,.
\eea
Thus, the $AA$ channel turns out to be $p$-wave suppressed, and so it is not relevant for indirect detection at present.
However, the $AA$ channel, if open kinematically, still contributes to the thermal annihilation cross section at the freeze-out era.

\section{Collider bounds and dark matter indirect detection
\label{sec:results}}

\subsection{Bounds from the LHC Run-1}

For a scalar with $m\sim 750$ GeV, there are stringent bounds from the LHC Run-1 
on various channels $pp\to A \to XX$, $\sigma(pp\to A \to XX)$, in terms of $\sigma(pp\to A) \times {\rm Br}(A\to XX) \equiv \sigma_A \times {\rm Br}_{XX}$:
\begin{itemize}
\item CMS \cite{Khachatryan:2015qba} ``Search for diphoton resonances in the mass range from 150 to 850 GeV in pp collisions at $\sqrt{s} =$ 8 TeV": $\sigma_A \times {\rm Br}_{\gamma\gamma} \lsim 2~ {\rm fb}$.
\item ATLAS \cite{Aad:2014fha} ``Search for new resonances in $W\gamma$ and $Z\gamma$ final states in $pp$ collisions at $\sqrt s=8$ TeV with the ATLAS detector'':  $\sigma_A \times {\rm Br}_{Z \gamma} \lsim 6 ~{\rm fb}$.
%
\item CMS \cite{Aad:2014aqa}: $\sigma_A \times {\rm Br}_{jj} \lsim 2.5 $ pb.
\item CMS \cite{Khachatryan:2014rra} ``Search for dark matter, extra dimensions, and unparticles in monojet events in proton-proton collisions at $\sqrt{s} = 8$ TeV'' and ATLAS \cite{Aad:2015zva} ``Search for new phenomena in final states with an energetic jet and large missing transverse momentum in pp collisions at $\sqrt{s}=$8 TeV with the ATLAS detector'': $\sigma(\slashed{E}_T +jet) \lsim 6 ~{\rm fb}$ for $\slashed{E}_T>500$ GeV.
\end{itemize}

\subsection{New data from the LHC Run-2}

The recently announced data from the Run-2 with $\sqrt{s}=13$ TeV include the hint of new resonance at 750 GeV and
the extracted cross section and the width for the $\sim750$ GeV resonance particle are:
\begin{itemize}
\item $\sigma(pp \to A)_{13 \rm TeV} \times {\rm Br}(A\to \gamma\gamma) =  (4.4\pm 1.1)~{\rm fb}$~\cite{Buttazzo:2015txu},
\item $\Gamma_{\rm tot} \approx 45$ GeV or $\Gamma_{\rm tot}/M \approx 6~\%$~\cite{Franceschini:2015kwy}.\footnote{A narrow decay width $\Gamma_{\rm tot}\sim 7$ GeV is discussed in Ref.~\cite{McDermott:2015sck} and also Ref.~\cite{Falkowski}.}
\end{itemize}
We also note that no resonances at $\sim750$ GeV are seen in $WW, ZZ, \ell^+ \ell^-$, or $jj$ events.\footnote{For non-standard interpretations of the 750 GeV diphoton excess, see e.g. Ref.~\cite{Cho:2015nxy}.}

%
\begin{figure}
\begin{center}
\includegraphics[width=0.49\linewidth]{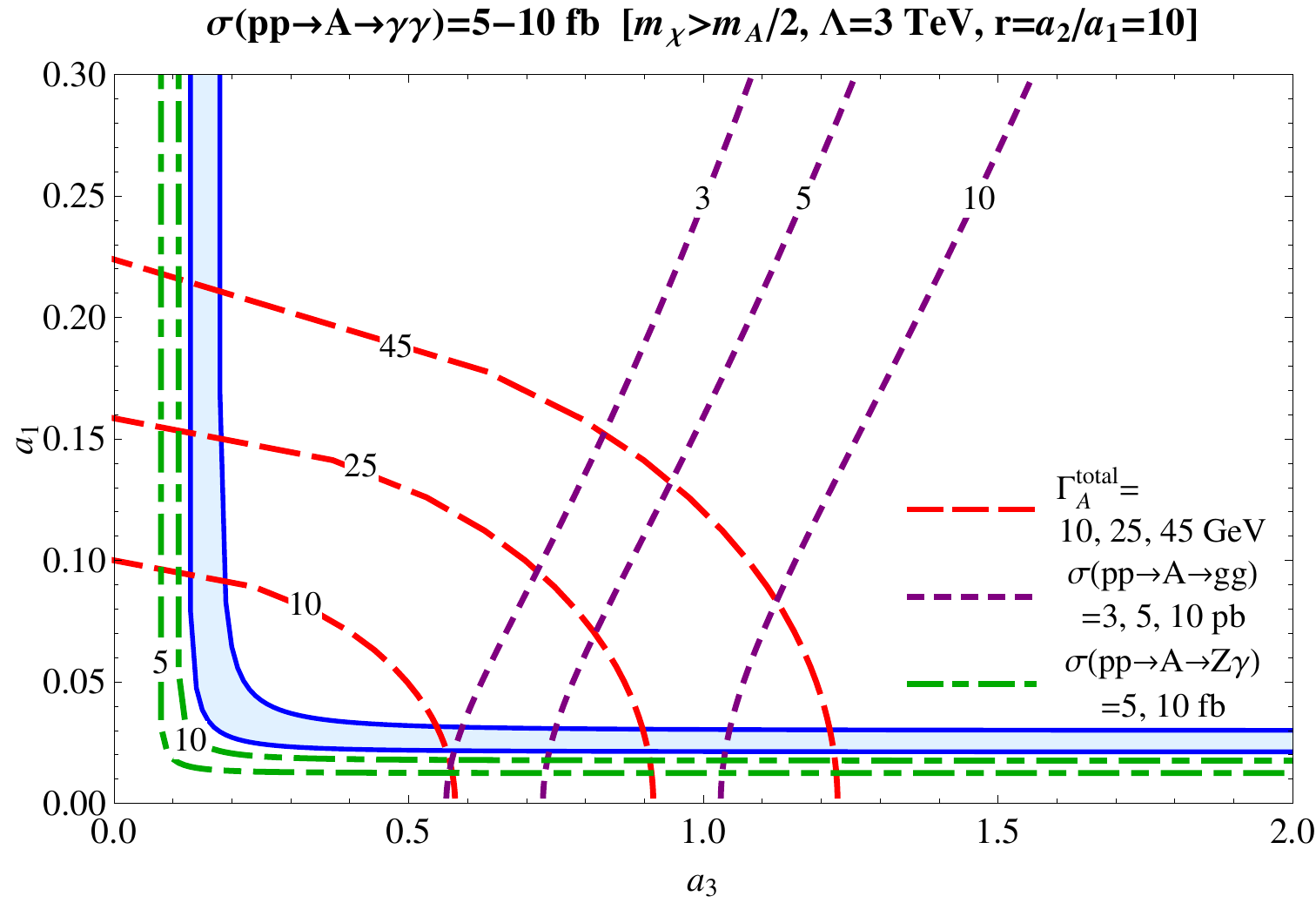}
\hspace*{0.1cm}
\includegraphics[width=0.49\linewidth]{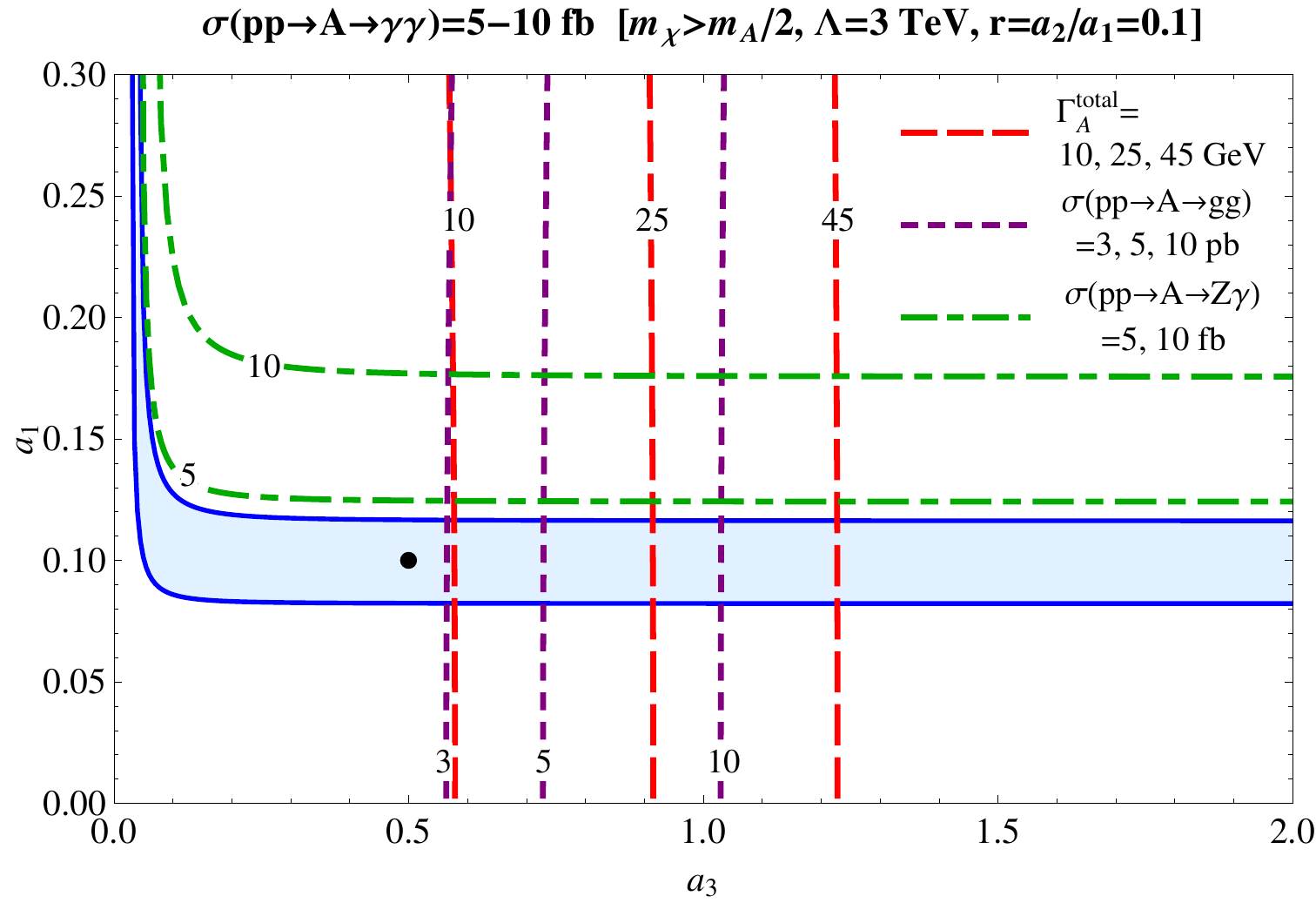}\\
\vspace*{0.3cm}
\includegraphics[width=0.49\linewidth]{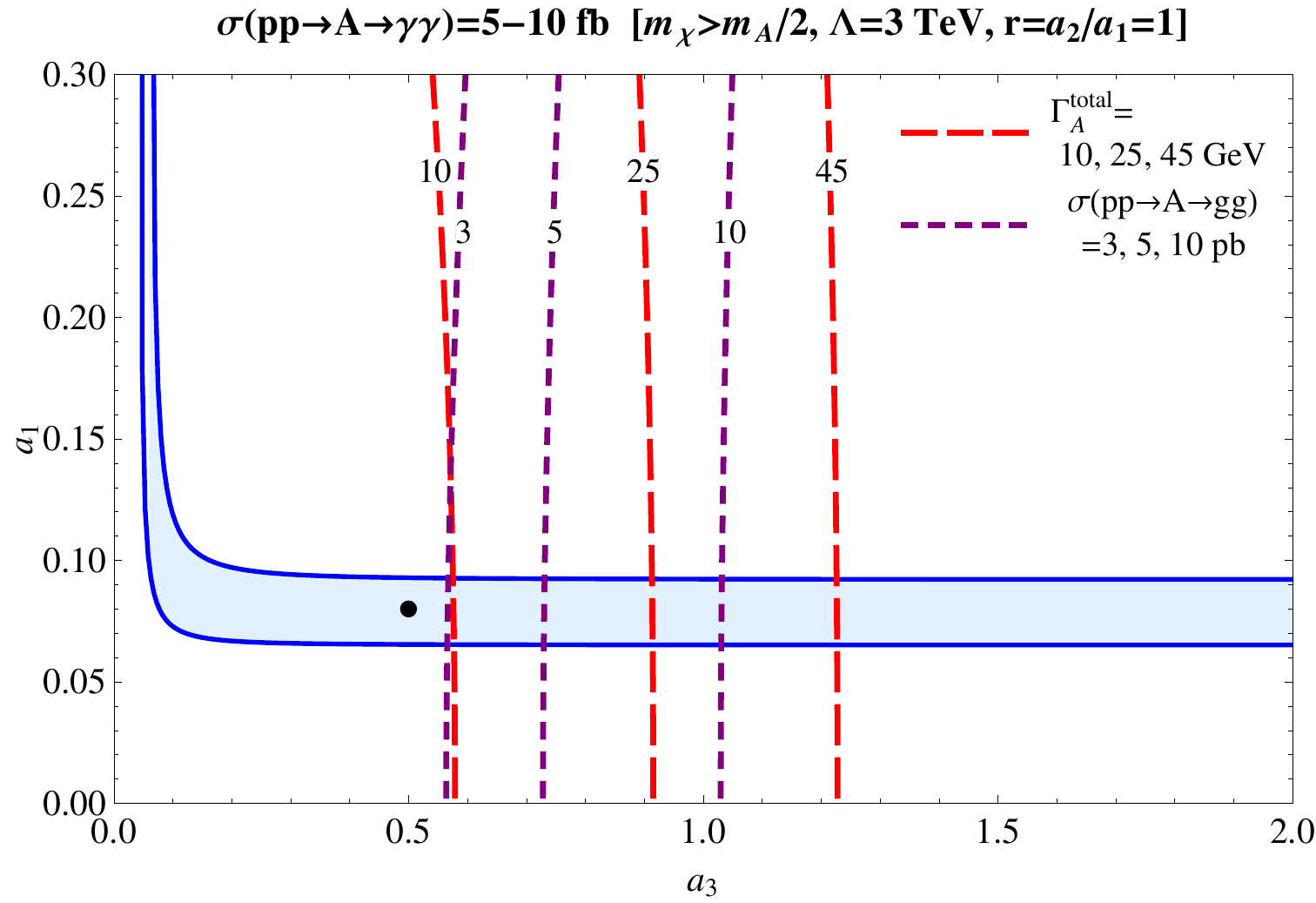}
\hspace*{0.1cm}
\end{center}
\vspace*{-0.7cm}
\caption{
Allowed parameter space in the $a_3-a_1$ plane for `heavy DM' ($2m_\chi > m_A$) with three representative $r=a_2/a_1$ ratios of $r=10, 0.1$, and 1.
In the blue band, $\sigma(pp \to A \to \gamma\gamma) \equiv \sigma_A \times {\rm Br}_{\gamma\gamma}$ is in the range of 5--10 fb (bottom to top).
The red dashed, purple dotted, and green dot-dashed lines show the total decay width of $A$, $\sigma(pp \to A \to gg)$, and $\sigma(pp \to A \to Z\gamma)$, respectively.
The black dots represent the benchmark points which will be used for the DM analysis in the following section.}
\label{Fig-LHC-LargeM}
\end{figure}
%

%
\begin{figure}
\begin{center}
\includegraphics[width=0.49\linewidth]{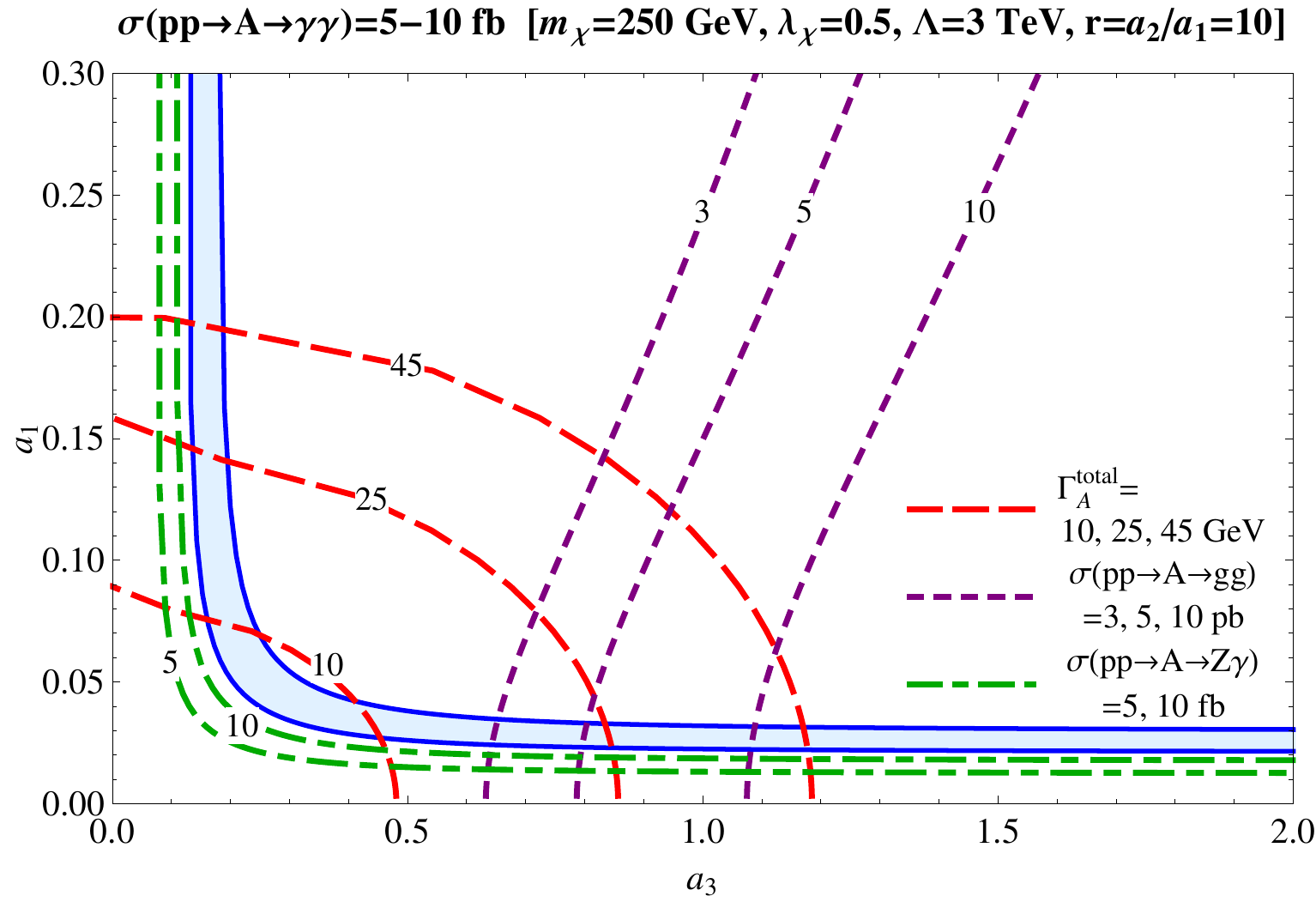}
\hspace*{0.1cm}
\includegraphics[width=0.49\linewidth]{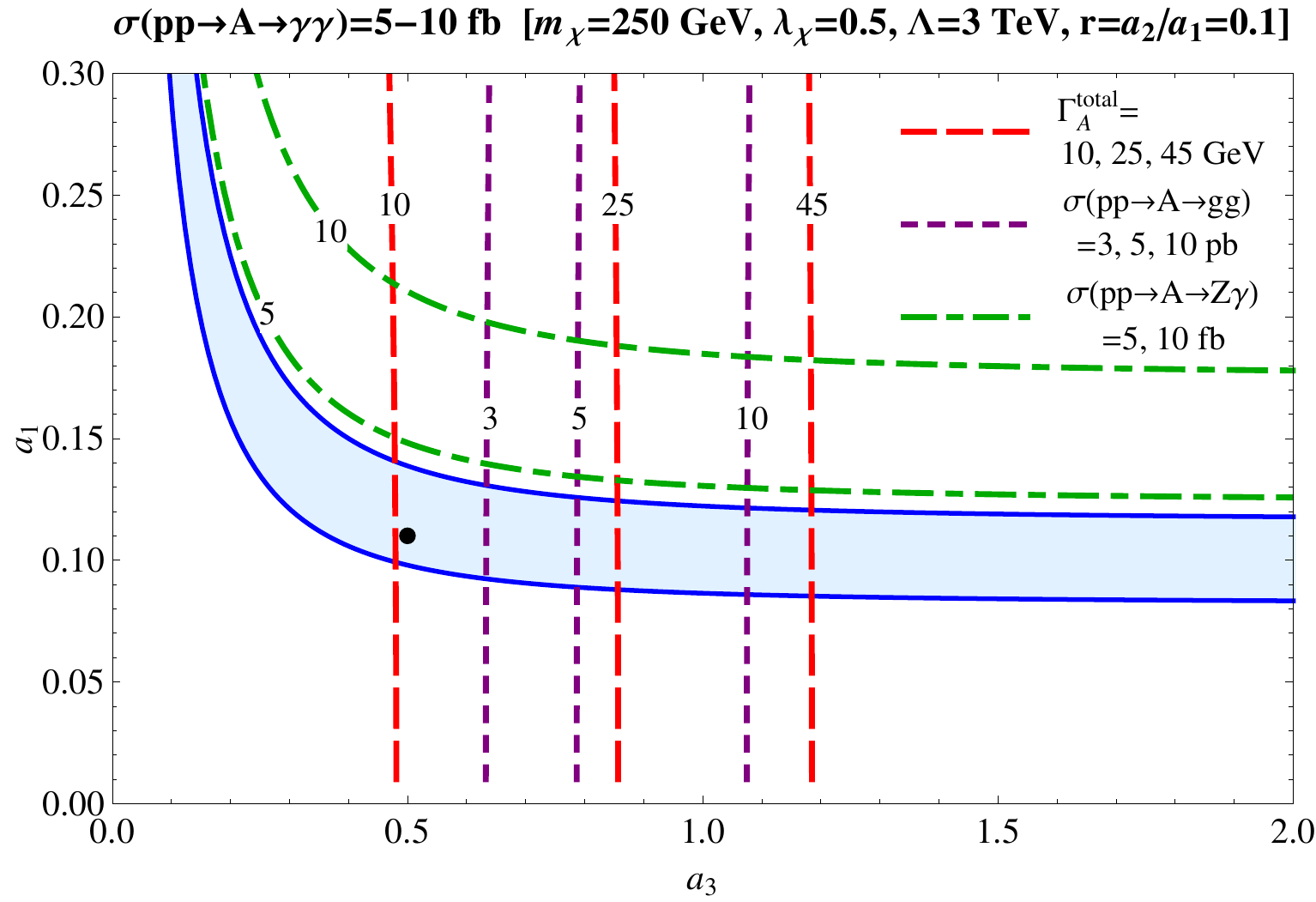}\\
\vspace*{0.3cm}
\includegraphics[width=0.49\linewidth]{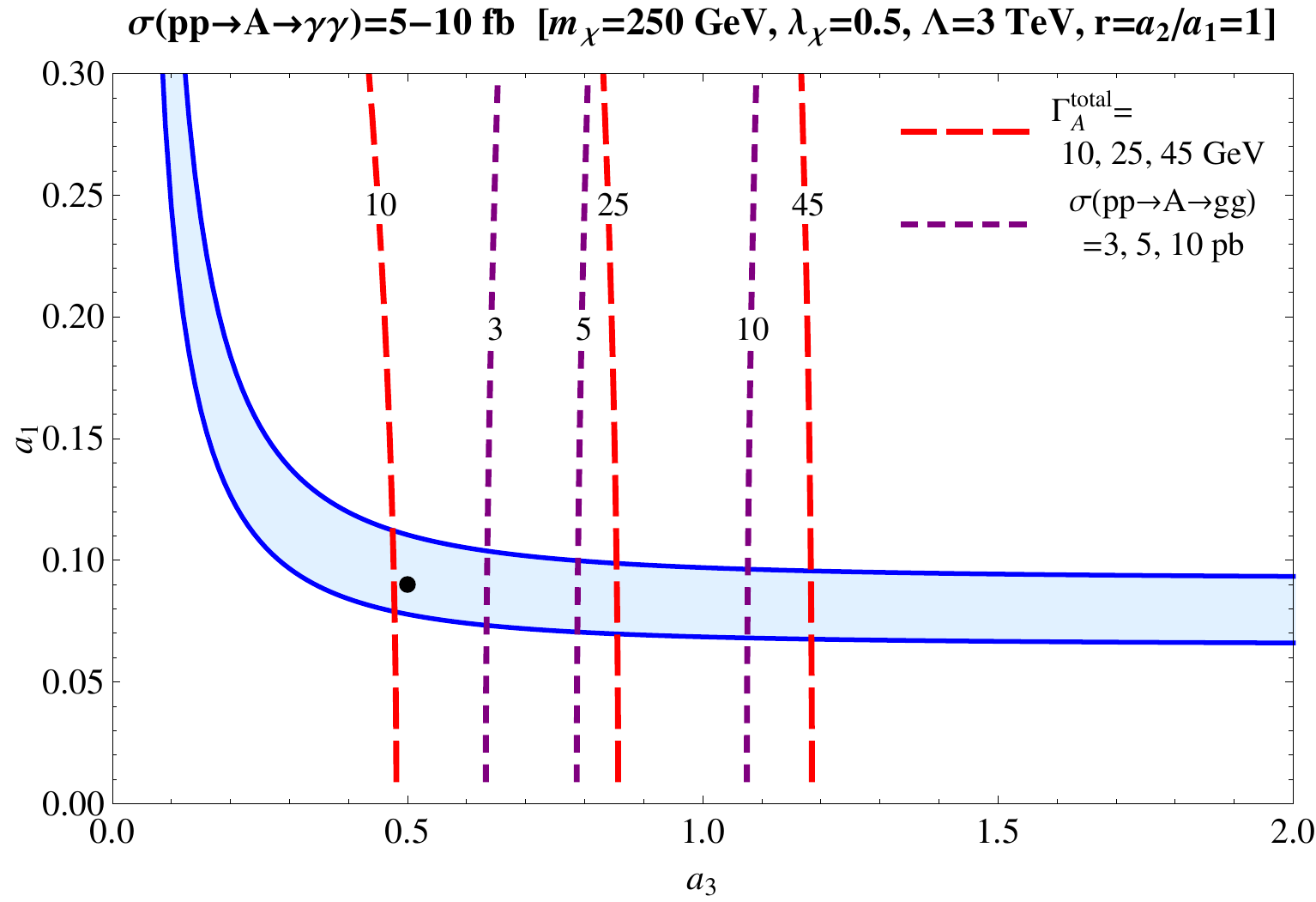}
\hspace*{0.1cm}
\end{center}
\vspace*{-0.7cm}
\caption{
Allowed parameter space for `light DM' ($2m_\chi < m_A$): a representative value of $m_\chi=250~{\rm GeV}$ is taken.
Each line is the same as Fig.~\ref{Fig-LHC-LargeM}.
}
\label{Fig-LHC-SmallM}
\end{figure}
%

The new pseudo-scalar particle $A$ can be produced via a gluon fusion process through the dimension five operator proportional to $a_3/\Lambda$, and thus $\sigma(pp \to A) \propto (a_3/\Lambda)^2$.
Without loss of generality, we take $\Lambda=3$ TeV by redefining $a_3$.
The coefficients $(a_1, a_2)$ can be redefined by a set of parameters $(a_1, r\equiv a_2/a_1)$.
We choose three representative values of $r=(10, 0.1, 1)$ for illustration, which represent the extreme cases: $a_2 \gg a_1$, $a_2 \ll a_1$, and $a_2 \approx a_1$, respectively.
The allowed parameter space in the $a_3-a_1$ plane will be shown for these representative cases as shown in Fig.~\ref{Fig-LHC-LargeM} ($2m_\chi > m_A$) and Fig.~\ref{Fig-LHC-SmallM} ($2m_\chi < m_A$).
For the `light DM' case with $2m_\chi < m_A$, the partial decay rate of the particle $A$ into a pair of DM particle, $\Gamma_A(\bar{\chi}\chi)$ in Eq.~(\ref{Gamma2DM}), decreases when $m_\chi$ approaches to $m_A/2\simeq375$ GeV due to the kinematic suppression factor where the results become very similar to those for $2m_\chi > m_A$ in Fig.~\ref{Fig-LHC-LargeM}.
On the other hand, for a very light DM ($m_\chi \ll m_A$), the partial decay rate saturates as $\Gamma_A(\bar{\chi}\chi) \approx \lambda_\chi^2 m_A/8\pi$.
Moreover, for smaller $m_\chi$, larger $\lambda_\chi$ is required to get the right DM relic abundance as can be seen from Fig~\ref{Fig-results-SmallM} in the next DM phenomenology section; therefore, $m_\chi$ cannot be a too small for a given value of $\lambda_\chi < {\cal O}(1)$ lying within the perturbative range.
For smaller $\lambda_\chi$, the results obviously come closer to those for $2m_\chi > m_A$.
Taking all these facts into account, we choose an intermediate mass $m_\chi=250$ GeV and relatively large coupling $\lambda_\chi=0.5$ as a benchmark point giving distinctive features from those in the `heavy DM' cases.

The blue band shows the region of $\sigma(pp \to A) \times {\rm Br}(A \to \gamma\gamma) =5-10$ fb with $\sqrt{s}=13$ TeV, which allows the fit to the resonance.
Here we allow a bit larger range of uncertainties than the cited value from e.g. Ref.~\cite{Buttazzo:2015txu} considering the slight gap between the CMS value ($\sim 760$ GeV) and the ATLAS value ($\sim 750$ GeV) of the resonance position.
The red dashed, purple dotted, and green dot-dashed lines represent the total decay width of $A$ (i.e., $\Gamma_A^{\rm total}$), $\sigma(pp \to A \to gg)$, and $\sigma(pp \to A \to Z\gamma)$, respectively.

Non observation of new resonance around 750 GeV in the $Z\gamma$ final state~\cite{Aad:2014fha} constrains larger $a_1$ region.
For $r=a_2/a_1 \gg 1$, one can easily find $a_{Z\gamma} > a_{\gamma\gamma}$ from Eqs.~(\ref{arr}) and (\ref{aZr}), and $\sigma(pp \to A \to Z\gamma)$ search result provides a stringent constraint on $\sigma(pp \to A \to \gamma\gamma)$ (see the upper-left panel of Figs.~\ref{Fig-LHC-LargeM} and \ref{Fig-LHC-SmallM}).
On the other hand, $a_{Z\gamma}=0$ for $r=1$, and thus we have no limit from $\sigma(pp \to A \to Z\gamma)$ data.
In addition, larger $a_3$ region is constrained by new resonance search in the $jj$ final state~\cite{Aad:2014aqa}
since $\sigma(pp \to A \to gg) \sim a_3^2$.
Therefore, for $r \lesssim \mathcal{O}(1)$, the blue band region with smaller $a_3$ is allowed by LHC observations.
Avoiding too narrow width of the diphoton resonance, we can finally choose four benchmark parameter sets: Scenario-I $\&$ II for $2m_\chi > m_A$ and Scenario-III $\&$ IV for $2m_\chi < m_A$ with a fixed $\Lambda$= 3 TeV which are depicted as black dots in Figs.~\ref{Fig-LHC-LargeM} and \ref{Fig-LHC-SmallM}.
The chosen benchmark parameter sets are:
\bea
\left.\begin{matrix}
\text{Scenario-I:}\quad a_1 &=& 0.1, ~a_3 = 0.5\quad {\rm for}\quad r=0.1 \label{bench1} \\
\text{Scenario-II:}\quad a_1 &=& 0.08, ~a_3 = 0.5\quad {\rm for}\quad r=1 \label{bench2}
\end{matrix}\right\} (\text{`heavy DM'}~2m_{\chi}>m_A),\\
\left.\begin{matrix}
\text{Scenario-III:}\quad a_1 &=& 0.11, ~a_3 = 0.5\quad {\rm for}\quad r=0.1 \label{bench3} \\
\text{Scenario-IV:}\quad a_1 &=& 0.09, ~a_3 = 0.5\quad {\rm for}\quad r=1 \label{bench4}
\end{matrix}\right\} (\text{`Light DM'}~2m_{\chi}<m_A).
\eea

Note that there exists no allowed region for $r=10$ (i.e. $r \gg 1$) due to the limit on $\sigma(pp \to A \to Z\gamma)$.

In Scenario-I and Scenario-III, $a_2 \ll a_1$ such that the relative branching fractions to various channels, ${\rm Br}(A\to XX)$ where $XX=\gamma\gamma, Z\gamma, ZZ, WW, gg$, are given as:
\bea
(\text{Scenario-I\&III})\quad\quad\quad
 &&\gamma\gamma: \gamma Z: ZZ: WW: gg \\
 \approx~ &&1 : 2 \tan^2\theta_W : \tan^4\theta_W : ~\ll1 : \frac{8}{\cos^4\theta_W} \left(\frac{a_3}{a_1}\right)^2\,.
\eea
On the other hand, in Scenario-II and Scenario-IV, $a_1=a_2$ such that the branching fractions are more democratically distributed except the vanishing in $Z\gamma$ channel:
\bea
(\text{Scenario-II\&IV})\quad\quad\quad
 &&\gamma\gamma: \gamma Z: ZZ: WW: gg \quad\quad\quad\quad\quad   \\
 \approx~ &&1 : 0 : 1 : 2 : 8 \left(\frac{a_3}{a_1}\right)^2\,. \quad\quad\quad\quad\quad
\eea

The branching fractions for Scenario-III and IV are almost same as Scenario-I and II except for the additional branching fraction to the dark matter which strongly depends on the mass $m_\chi$ and the coupling constant $\lambda_\chi$.
These specific patterns should be regarded as important predictions of each scenario that will be tested with more data in coming years.
In the next section, we will present our results for the `heavy DM' with $2m_\chi > m_A$ and `light DM' with $2m_\chi < m_A$ one after the other.

\subsection{Dark matter indirect detection
\label{sec:dm}}

The identified parameter space from the LHC data would provide a valuable information about the dark matter physics.
Using the four Scenarios (Scenario-I -- IV) obtained by the LHC data analysis,
we first calculate the relic density of dark matter $\chi$ and indirect detection limits from various cosmic-ray measurements and also the expected coverage by future experiments.\footnote{In Ref.~\cite{Chun:2015mka}, indirect detection limits from various cosmic-ray measurements are well summarized; thus, in this analysis we basically follow Ref.~\cite{Chun:2015mka}.}

%
\begin{figure}[t]
\begin{center}
\includegraphics[width=0.49\linewidth]{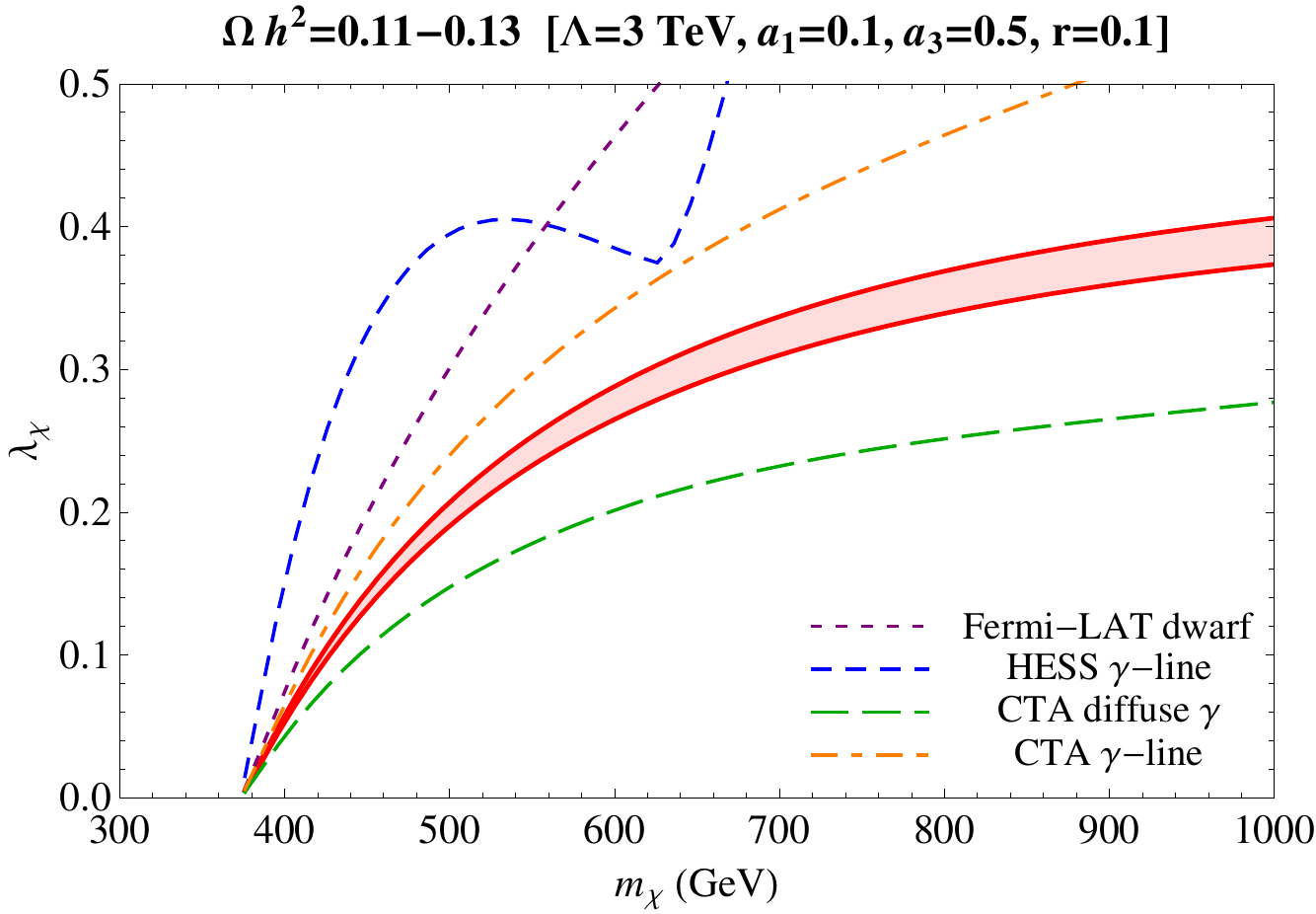}
\includegraphics[width=0.49\linewidth]{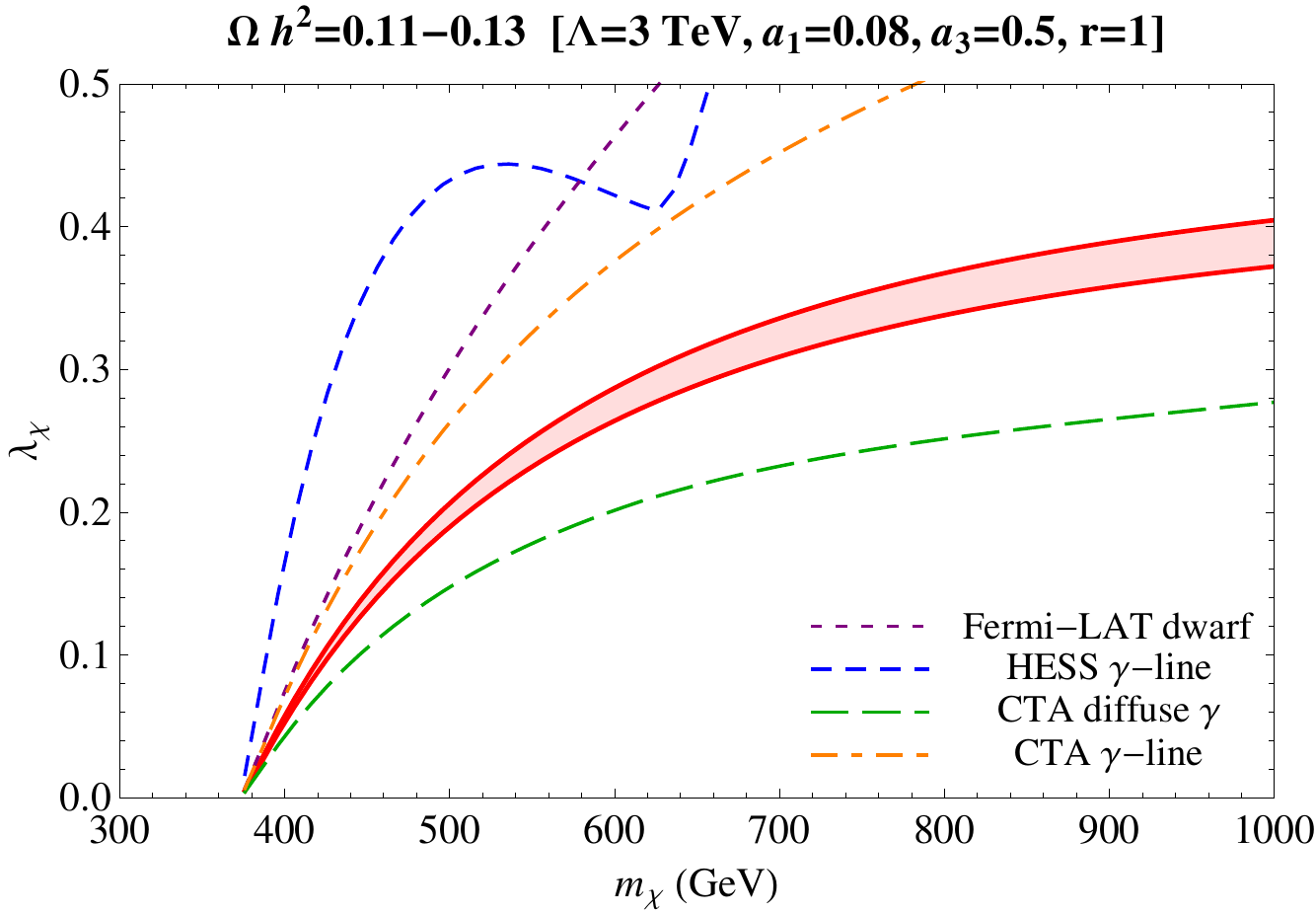}
\hspace*{0.1cm}
\end{center}
\vspace*{-0.7cm}
\caption{
Allowed parameter space in the $m_\chi-\lambda_\chi$ plane for `heavy DM' ($2m_\chi > m_A$) with two benchmark parameter sets given by Eq.~(\ref{bench1}).
In the red band, the relic density of DM $\chi$ is in the range of $0.11 < \Omega h^2 < 0.13$.
The purple dotted and blue dashed lines show the current upper limits obtained from the Fermi-LAT gamma-ray measurements from MW dwarf galaxies~\cite{Ackermann:2015zua} (the $gg$ channel) and the HESS line-like photon signature search~\cite{Abramowski:2013ax} (the $\gamma\gamma$ and $Z\gamma$ channels), respectively.
The CTA future sensitivities on the $gg$~\cite{Lefranc:2015pza, Carr:2015hta} and $\gamma X$~\cite{Bergstrom:2012vd, Ibarra:2015tya} channels are presented as green long-dashed and orange dot-dashed curves, respectively.}
\label{Fig-results-LargeM}
\end{figure}
%

%
\begin{figure}[t]
\begin{center}
\includegraphics[width=0.49\linewidth]{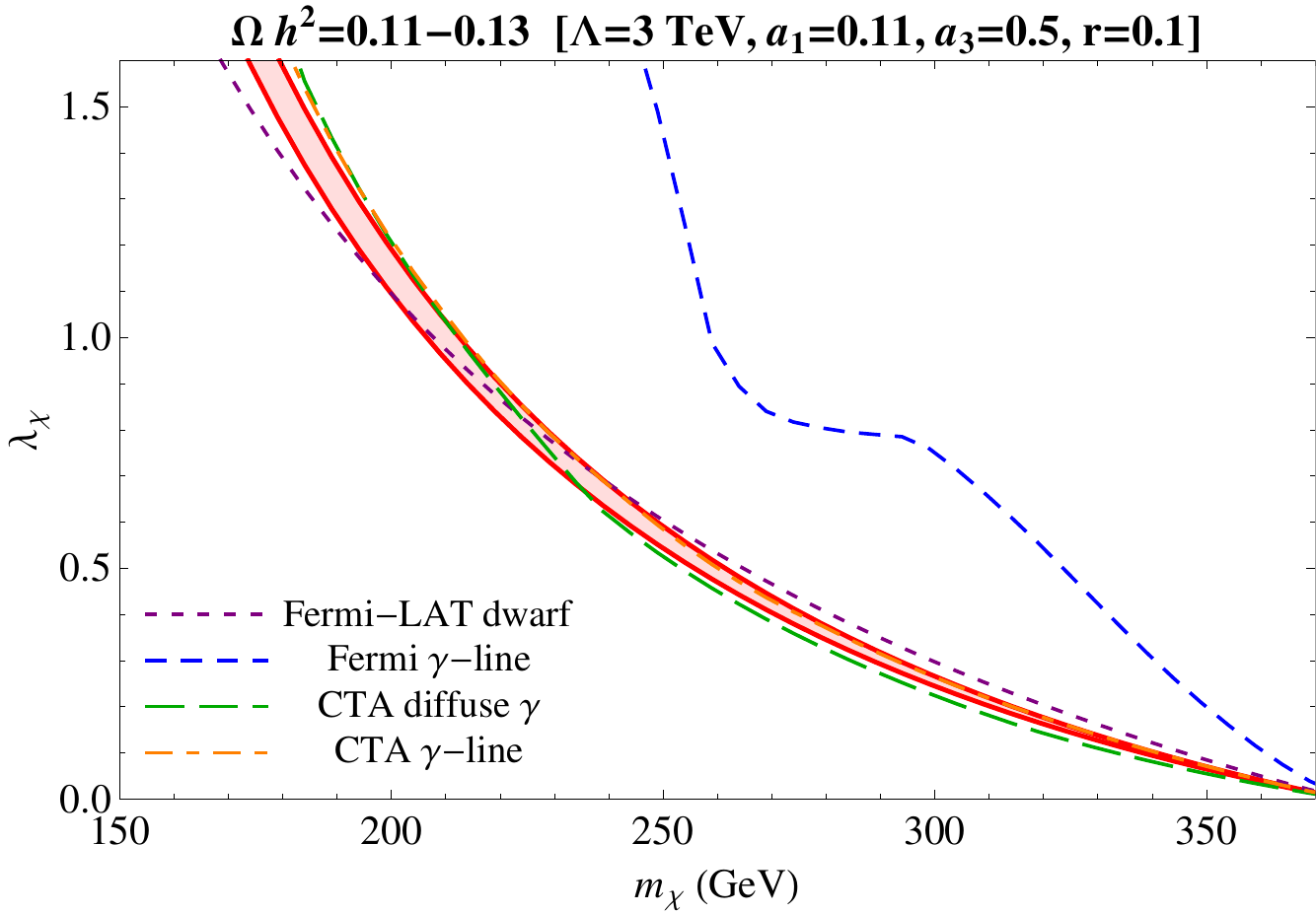}
\includegraphics[width=0.49\linewidth]{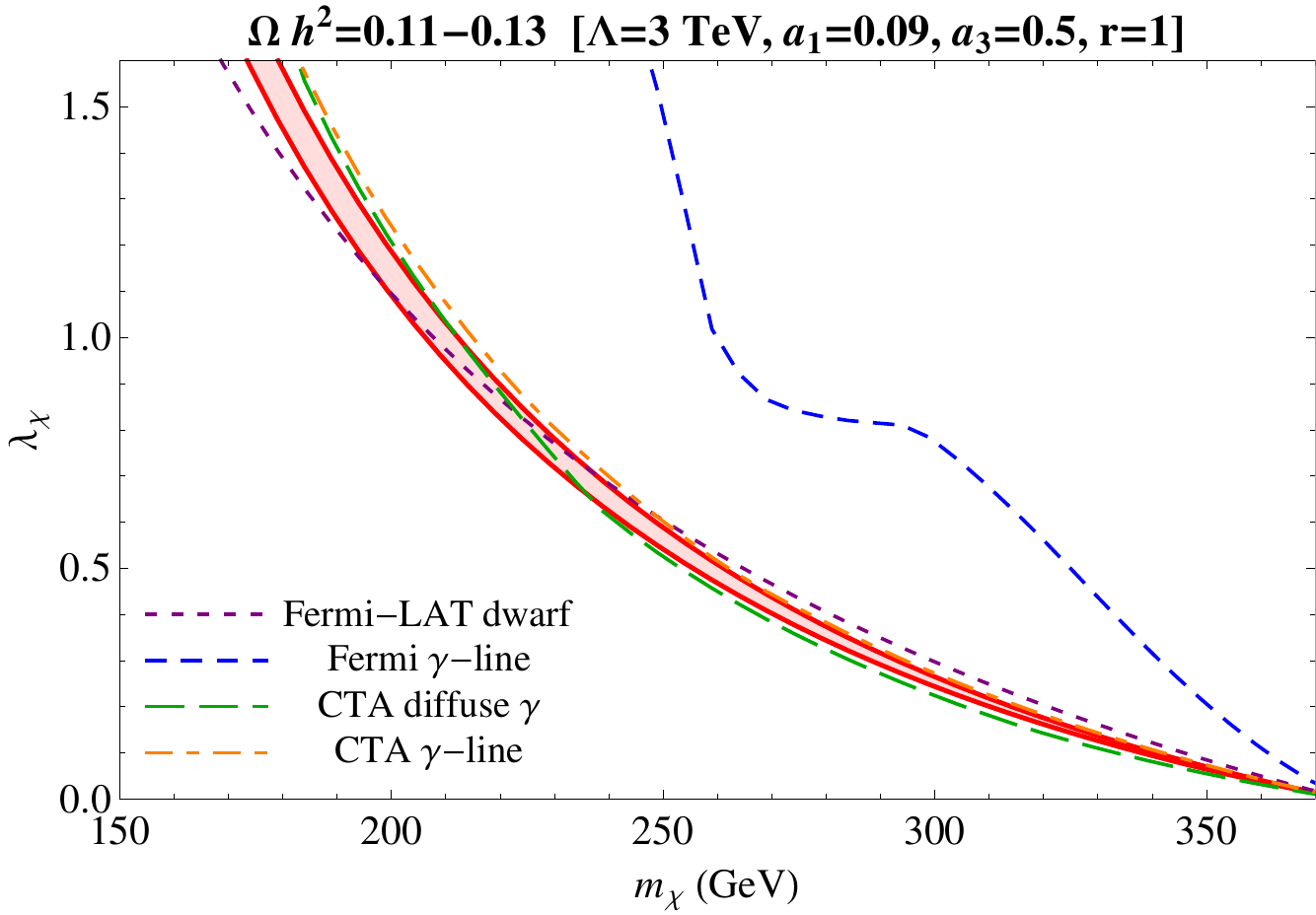}
\hspace*{0.1cm}
\end{center}
\vspace*{-0.7cm}
\caption{
Allowed parameter space for `light DM' ($2m_\chi < m_A$) with two benchmark parameter sets given by
Eq.~(\ref{bench3}).
Each line is the same as Fig.~\ref{Fig-results-LargeM}.
}
\label{Fig-results-SmallM}
\end{figure}
%

In Figs.~\ref{Fig-results-LargeM} and \ref{Fig-results-SmallM}, we present the combined results for preferred parameter regime with various observational constraints:
\begin{itemize}
\item The preferred parameter space for DM in the $m_\chi-\lambda_\chi$ plane is shown taking the DM relic abundance of $0.11 < \Omega h^2 < 0.13$~\cite{Ade:2015xua} depicted by the (red) colored band between two (red) solid lines.

\item The region below the band is not preferred in the standard thermal history because the interaction is too weak
    thus leads to overclosure of the universe with too large relic abundance, $\Omega_\chi \propto 1/\lambda_\chi^2$.
    On the other hand, above the band the amount of relic abundance is too small to explain the dark matter amount in the universe so that one needs to introduce additional source of dark matter beyond the current discussion.

\item     The upper regions of the purple dotted and blue dashed lines are excluded by the Fermi-LAT gamma-ray measurements from Milky Way (MW) dwarf spheroidal satellite galaxies~\cite{Ackermann:2015zua} (the $gg$ channel) and the HESS~\cite{Abramowski:2013ax} (for `heavy' DM) or Fermi-LAT~\cite{Ackermann:2015lka} (for `light' DM) line-like photon signature search from around the Galactic Center (GC) (the $\gamma\gamma$ and $Z\gamma$ channels) at $2\sigma$ level, respectively.
    In Ref.~\cite{Ackermann:2015zua}, the bound is derived just assuming the DM distribution in dwarf galaxies follows the Navarro-Frenk-White (NFW) profile~\cite{NFW}
    since the total mass of a dwarf galaxy within the half-light radius and the integrated J-factor have been found to be quite insensitive to the used DM density profile~\cite{Martinez:2009jh, Strigari:2013iaa, Ackermann:2013yva}.
    In Ref.~\cite{Ackermann:2015lka}, it has been shown that the limit is just 2 -- 4 times weaker even for the isothermal profile than for the Einasto profile~\cite{Graham:2005xx}
    because the Fermi-LAT has observed gamma-rays from all the sky and therefore can find the corresponding optimized regions of interest (ROI) for each DM profile.
    On the other hand, the HESS has searched gamma-ray line signatures for the ROI of a $1^\circ$ radius circle around the GC; thus, the HESS bound for the Einasto profile can be reduced by about two orders of magnitude~\cite{Chun:2015mka} for a more cored profile such as the isothermal profile.
    In this work, we use the HESS/Fermi-LAT photon line search limits for the Einasto profile.
    Those limits are so strong that if any dark matter with stronger emission rate with a larger value of $\lambda_\chi$ is all excluded.
    Especially, for `light DM' ($2m_\chi < m_A$), the preferred parameter space with the right relic abundance (the red band) with large $\lambda_\chi$ ($\gtrsim 0.7-1.1$ depending on $m_\chi$) is already constrained by the current Fermi-LAT gamma-ray observation on the MW dwarf galaxies through the $gg$ channel.

\item The $WW$ and $ZZ$ channels are also constrained by the gamma-ray observations.
    However, the limits on the $WW$ and $ZZ$ channels are much weaker than that on the $gg$ channel, which we show, due to much smaller annihilation rates: for Scenario-I -- IV,
    $\langle \sigma v_{\rm rel}\rangle_{ZZ, WW} / \langle \sigma v_{\rm rel}\rangle_{gg} < 0.01$.

\item We also show the near future coverage by planned observation: the CTA future sensitivities following Ref.~\cite{Chun:2015mka, Chun:2016cnm} on the $gg$~\cite{Lefranc:2015pza, Carr:2015hta} and $\gamma X$~\cite{Bergstrom:2012vd, Ibarra:2015tya} channels as green long-dashed and orange dot-dashed curves, respectively.
    We use the CTA future sensitivities for the Einasto profile,
    which however can be also weakened by about two orders of magnitude~\cite{Carr:2015hta, Ibarra:2015tya} for a cored profile such as the Burkert profile~\cite{Burkert:1995yz}.
    Interestingly, for `heavy DM' ($2m_\chi > m_A$), all of preferred parameter space for the right relic abundance (the red band) would be covered by the future CTA gamma-ray observation on the Galactic Halo through the $gg$ channel;
    however, for `light DM' ($2m_\chi < m_A$), the narrow preferred region with $m_\chi \approx 200-235$ GeV might not be covered even by the future CTA observation.
    Thus, we would call for further experimental and observational effort in this direction.
\end{itemize}

\section{Conclusion}

The recently reported diphoton resonance at $\sim750$ GeV brings our attention to a new portal for interactions between the standard model and dark matter sectors.
By identifying the mediator particle as a pseudo-scalar or axion-like-particle with 750 GeV mass,
we study a phenomenological model for dark matter interactions and found the preferred parameter space from the collider experiments.
It is very interesting to notice that indeed there exists a good chance of detecting dark matter signatures in dark matter annihilation through the portal interaction, $\chi\chi \to A \to gg$ and also small chances in $\gamma\gamma/\gamma Z$ in near future.
Future experiments from both collider experiments as well as cosmic-ray detection experiments, especially the CTA gamma-ray observation, will shed more light on this new avenue.

\section*{Acknowledgments}
JCP is supported by the National Research Foundation of Korea (NRF-2013R1A1A2061561, 2016R1C1B2015225).
SCP was supported by the National Research Foundation of Korea (NRF) grant funded by the Korean government (MSIP) (No. 2016R1A2B2016112).

\end{document}